\documentclass[aps,pre,twocolumn,showpacs,floatfix]{revtex4} 
\usepackage{amsmath,amssymb,graphicx,times,bm}

\newcommand{\bw}{\begin{widetext}}
\newcommand{\ew}{\end{widetext}}
\newcommand{\kommentar}[1]{}
\newcommand{\pavg}{\Pi(t)}

\begin{document}
 
\title{Dissipative Dynamics with Trapping in Dimers 
}
\author{Oliver M{\"u}lken}
\author{Lothar M{\"u}hlbacher}
\author{Tobias Schmid}
\author{Alexander Blumen}
\affiliation{
Physikalisches Institut, Universit\"at Freiburg,                                                 
Hermann-Herder-Stra{\ss}e 3, 79104 Freiburg, Germany}

\date{\today} 
\begin{abstract} 
The trapping of excitations in systems coupled to an environment allows to
study the quantum to classical crossover by different means. We show how
to combine the phenomenological description by a non-hermitian
Liouville-von Neumann Equation (LvNE) approach with the numerically exact path
integral Monte-Carlo (PIMC) method, and exemplify our results for a
system of two coupled two-level systems. By varying the strength of the
coupling
to the environment we are able to estimate the parameter range in which
the LvNE approach yields satisfactory results. Moreover, by matching the
PIMC results with the LvNE calculations we have a powerful tool
to extrapolate the numerically exact PIMC method to long times.
\end{abstract}
\maketitle

\section{Introduction}

Recent years have seen growing interest in coherent energy transfer. For
instance, it has been pointed out that photosynthesis might benefit from
quantum mechanical features of the transfer of excitations created
by the incoming solar energy \cite{photosynthesis}. A series of papers has
modelled coherent dynamics in the light-harvesting system of the photosynthetic
unit showing that the coupling to an environment does not necessarily destroy
all coherent features -- even at room temperature -- but also can support the
coherent transfer of excitations \cite{light-harvesting}. The majority of these
studies use the Lindblad form of the Liouville von Neumann equation (LvNE) for
the reduced density operator of the system where the environmental degrees of
freedom have been traced out \cite{lindblad}. However, this approach is only
valid in a limited parameter range of the coupling to the environment.

In contrast, rapid experimental advances allow to manipulate and control
ultra-cold atoms to a large extent. This offers the possibility to study
coherent transport and the effect of environment changes
(e.g., an increase in the temperature). An ideal system to study the dynamics
of excitations are (frozen) Rydberg gases \cite{rydberg}, whose atoms can
have complex spatial arrangements, for which the coherent dynamics can be
efficiently modelled by continuous-time quantum walks
\cite{mb2008}. 

Moreover, it is possible to adjust the properties of specific
(``special'') single atoms such that the excitation gets to be absorbed at
these atoms \cite{muelken07,reetz-lamour}. In this way they mimic the
reaction center (RC) of photosynthesis, where the excitation gets trapped
and further processed. Both
systems can be viewed as being donor-acceptor units, where the excitation is
created at the donor (``normal'' Rydberg atoms, light-harvesting complex)
and gets absorbed at the acceptor (``special'' Rydberg atoms, RC). The
decay of the excitation at the acceptor allows to globally monitor the
transport dynamics.

While the LvNE approach allows for a phenomenological modelling, other methods
treat the system and the coupling to the environment in a microscopic
manner. Our method of choice is the path integral Monte Carlo (PIMC) technique,
see, e.g., \cite{optimized filter,LM_JCP_121}, which can be applied for
arbitrary system-environment coupling strengths. However, unlike the LvNE
approach, the real-time PIMC method is plagued by the notorious dynamical sign
problem~\cite{sign-problem}, which significantly limits the time scales
accessible to PIMC simulations. However, by combining the LvNE and the PIMC
methods we are able to study excitation dynamics and trapping over large
timescales and improve the numerical accuracy of our results.

Our model system is a dimer, represented by two coupled two-level systems
(TLS), one of which acting as trap, the dimer being coupled to
the environment. By assuming not too
strong couplings to the environment and a single initial excitation of one of the
TLS, one can map the two TLS onto a single TLS, if, without the trap, the
probability of finding the excitation in the system is conserved
\cite{gm2006}. We note that various systems with, e.g., radial symmetry
and a trap in the center can effectively be mapped onto the dimer if
initially the excitation is homogeneously distributed over the periphery
\cite{mbb2006}.

\section{Coherent exciton trapping}

In general, we consider the Hamiltonian of a network of $N$ nodes, where
each node represents a single two-level system. Let ${\bm H}_0$ be the
Hamiltonian without traps. The accessible Hilbert space is completely
characterized by the basis states $|j\rangle$ which are associated with
excitations localized at the nodes $j=1,\dots,N$. Within a
phenomenological approach, the Hamiltonian, which incorporates trapping of
excitations at the nodes $m \in \mathcal{M}$, $\mathcal{M} \subset \{1,
\dots, N\}$, is given by ${\bm H} \equiv
{\bm H}_0 - i{\bm \Gamma}$, where $i{\bm \Gamma} \equiv i \Gamma \sum_{m}
| m \rangle \langle m |$ is the trapping operator, see
Ref.~\cite{muelken07} for details. As a result, ${\bm H}$ is non-hermitian
and has $N$ complex eigenvalues, $E_l = \epsilon_l - i\gamma_l$
($l=1,\dots,N$) where $\gamma_l>0$, and $N$ right and $N$ left
eigenstates, denoted by $|\Phi_l\rangle$ and $\langle\tilde\Phi_l|$,
respectively. The transition probability from node $j$ to node $k$ is then given by
\begin{equation}
\pi_{k,j}(t)
= \Big|\sum_l \exp(-\gamma_lt) \exp(-i\epsilon_lt)\langle k |
\Phi_l \rangle \langle \tilde\Phi_l | j \rangle\Big|^2,
\end{equation}
so that the negative imaginary parts $\gamma_l$ of $E_l$ determine the temporal
decay. The mean survival probability $\pavg$ of an excitation in the presence
of $M$ trap nodes is a global property
of the network and is defined as
\begin{equation}
\pavg \equiv \frac{1}{N-M} \sum_{j\neq m}
\sum_{k\neq m}
\pi_{k,j}(t),
\end{equation}
i.e., $\pavg$ is the average of $\pi_{kj}(t)$ over all initial nodes $j$
and all final nodes $k$, neither of them being a trap node. 

\subsection{Liouville-von Neumann equation}

The Schr\"odinger equation can be recast into the Liouville-von Neumann
equation (LvNE) when considering the density operator ${\bm\rho}$. For
Hermitian Hamiltonians ${\bm H}_0$ one has $\dot{\bm \rho} = -i \big[ {\bm
H}_0, {\bm \rho} \big]$, where $[\cdot,\cdot]$ is the commutator. Now, for
the non-hermitian Hamiltonian ${\bm H}$ one obtains
\begin{equation}
\dot{\bm \rho} = -i \big[ {\bm H_0}, {\bm \rho} \big] -
\big\{{\bm\Gamma},{\bm\rho}\big\},
\end{equation}
where $\{\cdot,\cdot\}$ is the anti-commutator.

Introducing the coupling to the environment naturally complicates the
situation. However, under certain conditions one can employ the so-called
Lindblad form of the LvNE, where the Lindblad operators specify the
coupling \cite{lindblad}. Consider now Lindblad operators which can
be written as $\sqrt{\lambda_j}\bm L_j$, where $\lambda_j$ is a fixed
decay rate.  Assuming all rates to be equal, $\lambda_j\equiv\lambda$ for
all $j$, the LvNE reads
\begin{equation}
\dot{\bm \rho} = -i \big[ {\bm H_0}, {\bm \rho} \big] - 
\big\{{\bm\Gamma},{\bm\rho}\big\} - 2\lambda 
\sum_j \Big(\bm\rho - \langle j|\bm\rho| j\rangle\Big){\bm L}_j.
\label{lvne}
\end{equation}
The rate $\lambda$ can be estimated  from the spectral density
$J(\omega)$ describing the environment within the Caldeira-Leggett
model~\cite{caldeira_leggett__Ann.Phys._(1983-4)} at a given temperature
$T$. Taking $J(\omega) = 2\pi \alpha \omega \exp(-\omega/\omega_c)$ and
using the Markov approximation one arrives at $\lambda = \pi \alpha k_B T$
\cite{lindblad}. One has to bear in mind that Eq.~(\ref{lvne}) is an
approximation with a limited range of validity: For a very large coupling
strength $\lambda$, Eq.~(\ref{lvne}) leads to the quantum Zeno limit
rather than to a classical master/rate equation.  In the following we will
consider Lindblad operators which are given by projection operators of the
type ${\bm L}_j \equiv |j\rangle\langle j|$ \cite{lindblad}. 
  
\section{Dimers with Traps}

In the sequel we will consider a dimer which is coupled to an external
bath. This system allows to solve Eq.~(\ref{lvne}) analytically and,
moreover, to compare the approximate LvNE results to the numerically exact
PIMC calculations. The Hamiltonian of the dimer without any coupling to
the surroundings can be expressed through the Pauli matrices
${\bm\sigma_z}$ and ${\bm\sigma_x}$,
\begin{equation}
{\bm H} = E \ {\bm 1} - V{\bm\sigma_x} + i\frac{\Gamma}{2}({\bm 1} -
{\bm\sigma_z}) \,, 
\end{equation}
where $E$ is the onsite energy, which we choose to be the same for both
nodes, and $V$ is the coupling between the two nodes. It is easily
verified that the eigenvalues are
\begin{equation}
E_\pm= E \pm V e^{\pm i\phi} = E \pm \sqrt{V^2-\Gamma^2/4} - i \Gamma/2,
\end{equation}
where $\phi = \arcsin(\Gamma/2V)$. For $\Gamma\to0$ ($\phi\to0$) this
yields the correct eigenvalues $E\pm V$ of ${\bm H}_0$. Note that for
$\Gamma\leq2V$ the negative imaginary part of $E_\pm$ is identical for
both eigenvalues, i.e., $\gamma_+=\gamma_-=\Gamma/2$. The
bi-orthonormalized eigenstates of ${\bm H}$ are of the form
\begin{equation}
|\Phi_\pm\rangle \equiv \frac{1}{\sqrt{2\cos\phi}} \left( \begin{matrix}
e^{\pm i\phi/2} \\
\pm e^{\mp i\phi/2} \end{matrix} \right) 
\end{equation}
and
\begin{equation}
|\tilde\Phi_\pm\rangle \equiv \frac{1}{\sqrt{2\cos\phi}} \left(
\begin{matrix} e^{\mp i\phi/2} \\
\pm e^{\pm i\phi/2} \end{matrix} \right),
\end{equation}
where the phases $\phi$ depend on $\Gamma$ such that in the limit
$\Gamma\to0$ one recovers the eigenstates of ${\bm H}_0$.

We note, however, that finding the bi-orthonormal basis set is not necessary
for the following calculations. One just has to require that the basis
sets of $\bm H$ and $\bm H^\dagger$ are orthonormal, respectively. In this
way one diagonalizes $\bm H$ and $\bm H^\dagger$ separately, which in the
end leads to the same eigenvalues and eigenstates as the approach described above.

When the coupling to the environment vanishes ($ \lambda\to 0$) one obtains the
survival probability directly from the eigenstates and eigenvalues of $\bm
H$. For $\Gamma\leq2V$ one has
\begin{equation}
\Pi(t) = e^{-\Gamma t}\frac{\cos^2(\phi +
tV\cos\phi)}{\cos^2\phi} \quad (\mbox{for} \ \lambda=0).
\label{pi_l0}
\end{equation}
We note that for values $\Gamma> 2V$ the dimer is overdamped.
%


When considering the dimer without traps ($\Gamma=0$) but coupled to the
environment, Eq.~(\ref{lvne}) simplifies and, from the solution for ${\bm
\rho}$, one obtains the transition probabilities
\begin{eqnarray} 
\pi_{1,1}^{(0)}(t) = \frac{1}{2} &+& \frac{e^{-\lambda t}}{2} \Bigg[
\frac{\lambda\sin\left(t\sqrt{4V^2-\lambda^2}\right)}{\sqrt{4V^2-\lambda^2}}
\nonumber \\
&& +
\cos\left(t\sqrt{4V^2-\lambda^2}\right)
\Bigg] \quad (\mbox{for} \ \Gamma=0) \ \  \
\label{pi_g0}
\end{eqnarray}
and $\pi_{2,1}^{(0)}(t) = 1 - \pi_{1,1}^{(0)}(t)$. For $\lambda\to0$ one recovers the
simple oscillatory behavior of the transition probabilities [namely,
$\lim_{\lambda\to 0}\pi_{1,1}^{(0)}(t)=\cos^2(Vt)$]. For $\lambda>0$, i.e., with coupling to the
surroundings, the transition probabilities still show oscillations
superimposed on an exponential decay in time which tends to the classical
equipartition value of $1/2$. 


In order to combine the results of Eq.~(\ref{pi_l0}) -- we will only focus on
values $\Gamma <2V$ at this stage -- and Eq.~(\ref{pi_g0}), we expand in both
equations all terms except the exponentials to first order in $\Gamma$ and
$\lambda$, respectively. Note that for Eq.~(\ref{pi_l0}) we obtain a
product of $\exp(-\Gamma t)$ and $\cos^2Vt$, which is the simple
oscillatory behavior of the dimer without trap. Now, the coupling to the
environment affects all transitions but still conserves probabilities.
Therefore, we replace the term $\cos^2Vt$ by the expansion of
Eq.~(\ref{pi_g0}), by which we obtain
\begin{eqnarray}
\Pi(t) &\approx& e^{-\Gamma t} \pi_{1,1}^{(0)}(t)  \nonumber \\
&\approx&  e^{-\Gamma t} \Big[ \frac{1}{2} +\frac{e^{-\lambda t}}{2}
\Big(\cos 2Vt +\frac{\lambda}{2V}\sin 2Vt \Big) \Big].
\label{pi_lr_approx}
\end{eqnarray}

\subsection{PIMC}

In order to corroborate our results we will compare the phenomenological LvNE
approach described above to the numerically exact PIMC calculations based on a
microscopic modelling of the dissipative environment. This allows to judge for
which parameter range the approximation by Eq.~(\ref{lvne}) delivers
satisfactory results.

When coupling the dimer to a bath, the total Hamiltonian reads $\bm
H_\text{tot} = \bm H + \bm H_\text{I} + \bm H_\text{B}$, where the dimer-bath
coupling and bath are described in the framework of the Caldeira-Leggett model
\cite{caldeira_leggett__Ann.Phys._(1983-4)},
\begin{equation} \label{Caldeira-Leggett}
\bm H_\text{I} + \bm H_\text{B}
=
-{\bm\sigma_z} \sum_\kappa c_\kappa {\bm X_\kappa} + \sum_\kappa \left(
\frac{\bm P_\kappa^2}{2m_\kappa} + \frac{1}{2}m_\kappa\omega_\kappa^2 {\bm
X_\kappa^2} \right) \,.
\end{equation}
Here, $\bm P_\kappa$ and $\bm X_\kappa$ are the momentum and position
operators of the bath degrees of freedom, respectively, while $m_\kappa$
and $\omega_\kappa$ denote their mass and frequency.  The counter term,
which prevents a renormalization of the free dimer's energy levels due to
the environmental coupling \cite{weiss}, is absent in
Eq.~(\ref{Caldeira-Leggett}) since it reduces to a physically irrelevant
constant in the case of a two-level system. After tracing out the
environmental degrees of freedom, the onsite population of node
$|n\rangle$ becomes
\begin{equation} \label{pops_path}
\pi_{n,1}(t)
=
\oint\!{\cal D}\tilde{\sigma}\,
\delta_{\tilde{\sigma}(t),n} \exp\left\{
\frac{i}{\hbar}S[\tilde{\sigma}] - \Phi[\tilde{\sigma}] \right\} \,,
\end{equation}
where $\tilde{\sigma}$ denotes a closed quantum path in terms of the
eigenstates of ${\bm\sigma_z}$ with $\tilde{\sigma}(0) = -1$ (referring
to the initial preparation in node $|1\rangle$), and $S[\tilde{\sigma}]$ is the
action of the free dimer. The influence of the environment is summarized in the
Feynman-Vernon influence functional $\Phi[\sigma]$ \cite{feynman},
which is completely determined by the environment's spectral density,
\begin{equation} \label{spectral density}
J(\omega)
=
\frac{\pi}{2\hbar} \sum_\alpha \frac{c_\kappa^2}{m_\kappa\omega_\kappa}
\delta(\omega-\omega_\kappa) \,;
\end{equation}
for further details, we refer to Ref.~\cite{weiss}.

As the exact dynamics Eq.~(\ref{pops_path}) can not be calculated analytically,
one has to resort to a numerical evaluation of the path integral. Here, the
PIMC method has proven to be a promising approach to obtain numerically exact
results even in regions of parameter space where approximative methods fail
(for details, see~e.g.~Refs.~\cite{optimized filter,LM_JCP_121}). In our case
it is straightforward to adopt the approach presented in Ref.~\cite{LM_JCP_121}
once the free dimer's forward and backward propagators, which define
$S[\tilde{\sigma}]$, are expressed according to
\begin{eqnarray} 
\langle n | \exp(-i {\bm H}t/\hbar) | n'\rangle
&=&
\sum_{\sigma = \pm} \langle n | \Phi_\sigma\rangle\!\langle\tilde{\Phi}_\sigma
| n' \rangle\,
 e^{-i E_\sigma t/\hbar} \,,
\nonumber\\
\langle n' | \exp(i {\bm H^\dagger}t/\hbar) | n\rangle
&=&
\sum_{\sigma = \pm} \langle n' | \Phi_\sigma\rangle\!\langle\tilde{\Phi}_\sigma
| n \rangle\,
 e^{i E_\sigma^* t/\hbar} \,,
\nonumber\\
\end{eqnarray}
which, even for $\Gamma \neq 0$, are complex conjugate to each other.

\section{Comparison of LvNE to PIMC}

\begin{figure}[htb]
\centerline{\includegraphics[clip=,width=\columnwidth]{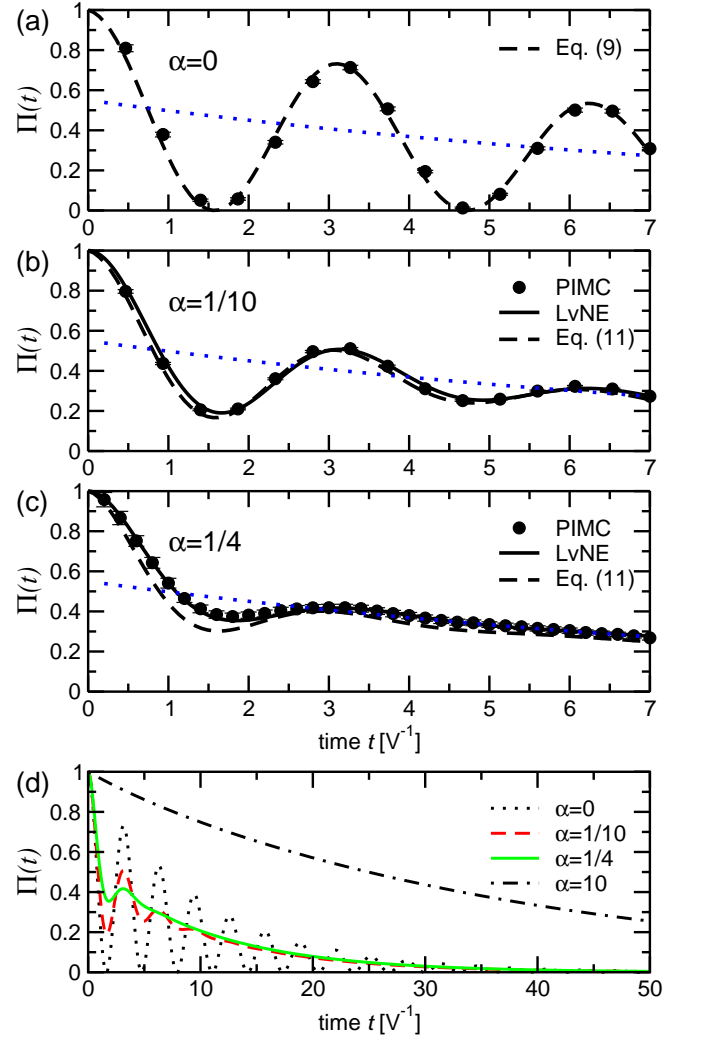}}
\caption{(Color online) PIMC results (circles) for a dimer with $\Gamma=0.1$
and different system-bath couplings $\alpha=\lambda/\pi$: (a) $\alpha=0$,
(b) $\alpha=1/10$, and (c) $\alpha=1/4$. The solid lines represent
the numerical solution of the LvNE equation, the dashed lines show the
corresponding analytical results obtained from Eq.~(\ref{pi_l0}) for
$\alpha=\lambda=0$ and from Eq.~(\ref{pi_lr_approx}) for $\alpha=1/10$ and
$\alpha=1/4$. The dotted blue line shows the long time limit $\Pi(t) \sim
\exp(-\Gamma t)$. Panel (d) shows the corresponding long-time behavior of
the numerical LvNE solution for the three different values of $\alpha$ and
additionally the behavior for large couplings $\alpha=10$.}
\label{pimc_dimer}
\end{figure}

Figure \ref{pimc_dimer} compares the survival probabilities of a dissipative
dimer obtained from the approximative LvNE approach to the numerically exact PIMC
calculations for a bath with ohmic spectral density~(\ref{spectral density})
with exponential cutoff, $J(\omega) = 2 \pi \alpha \omega
e^{-\omega/\omega_c}$. Here, the onsite energies $E$ and the coupling elements
$V$ have been taken to be equal, $E=V=1$, while the temperature is fixed
to $k_B T =V$, and we set $\omega_c=5V$. 

For small trapping strength ($\Gamma=0.1$) and vanishing coupling to the
environment ($\alpha=0$), Fig.~\ref{pimc_dimer}(a), the PIMC calculations
coincide with the result of Eq.~(\ref{pi_l0}). A moderate increase of the
coupling ($\alpha=1/10$), Fig.~\ref{pimc_dimer}(b), still leads for
Eqs.~(\ref{lvne}) (solid lines) and (\ref{pi_lr_approx}) (dashed lines) to
results which are in excellent agreement with the findings of the PIMC
calculations (symbols).  When increasing the coupling further to
$\alpha=1/4$, Fig.~\ref{pimc_dimer}(c), however, the approximate solution,
Eq.~(\ref{pi_lr_approx}), begins to deviate from the LvNE and the PIMC
calculations, which are still in very good agreement. 
 
As the numerical effort of real-time PIMC simulations grows exponentially
with time, they can cover only short-to-intermediate timescales.
However, the agreement between the LvNE and the PIMC calculations in the weak
coupling regime allows to compensate for this shortcoming, by using at
longer times the LvNE
results, see Fig.~\ref{pimc_dimer}(d).

As mentioned earlier, strong couplings $\lambda$ in the LvNE lead to the Zeno
limit, and therefore disagree with the large-coupling/high-temperature
behavior of the PIMC formalism. Figure
~\ref{pimc_dimer}(d) shows $\pavg$ for $\alpha=10$ (dashed-dotted line),
which clearly deviates from the long-time behavior of the curves for small
$\alpha$-values. This corroborates the fact that the LvNE in the Lindblad
form only yields the correct long-time behavior for rather weak couplings
to the environment.

\begin{figure}[htb]
\centerline{\includegraphics[clip=,width=\columnwidth]{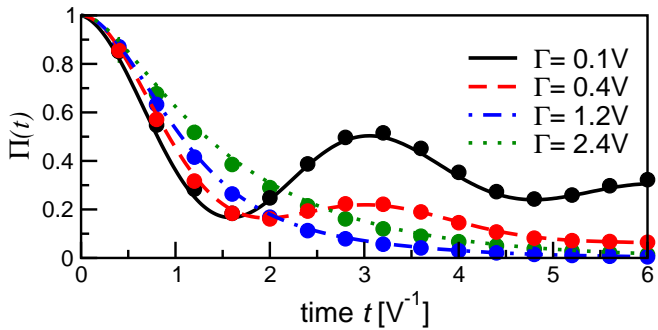}}
\caption{(Color online) 
Dimer with $\alpha=1/10$ and different trapping strengths $\Gamma$.  The
lines represent the numerical solution of the LvNE and the symbols show
the corresponding PIMC results.}
\label{pimc_dimer_trapping}
\end{figure}

For $\alpha = 1/10$ and varying $\Gamma$, Fig.~\ref{pimc_dimer_trapping}
compares $\pavg$ obtained from the numerical solution of the LvNE to the
results of the corresponding PIMC calculations. The trap is part of the
system, i.e., it directly enters the Hamiltonian $\bm H$. Therefore,
unlike a change in $\alpha$, a change in $\Gamma$ leads to no
significant differences for $\pavg$ obtained from the two methods.
Nevertheless, varying the trapping strength has strong implications for
the survival probability: At short times
($t<2$) the survival probability $\pavg$ is smaller for smaller $\Gamma$.
At larger times the decay is very pronounced for larger $\Gamma$-values;
smaller $\Gamma$-values do not dampen the oscillations which are
superimposed on the decay. This effect will be weakened when increasing
the coupling to the environment.

\section{The Incoherent Case}

Finally, in the limit of very strong coupling to the environment, all
coherences will be quickly destroyed and the resulting transport becomes purely
incoherent. The dynamics is governed by a master equation whose transfer
matrix is
\begin{equation}
{\bm T} = \left( \begin{matrix} \tilde E & -\tilde V \\ -\tilde V & \tilde
E + \tilde \Gamma \end{matrix}
\right),
\end{equation}
One obtains the parameters $\tilde E$, $\tilde V$, and $\tilde\Gamma$ by
fitting to the PIMC calculations at large temperatures. In principle one can
obtain the transfer rates $\tilde V$ from a golden rule approach
\cite{emw1994}.
The matrix $\bm T$ has purely real eigenvalues,
\begin{equation}
\lambda_\pm = \tilde E \pm \tilde V e^{\mp \psi}
= \tilde E +\tilde\Gamma/2 \pm \sqrt{\tilde V^2+\tilde \Gamma^2/4},
\end{equation}
Similar to the coherent case, the eigenstates of $\bm T$ read
\begin{equation}
|\Psi_\pm\rangle \equiv \frac{1}{\sqrt{2\cosh \psi}} \left( \begin{matrix}
e^{\pm \psi/2} \\
\pm e^{\mp \psi/2} \end{matrix} \right) 
\end{equation}
where $\psi = \text{arcsinh}(\tilde\Gamma/2\tilde V)$. Now, the
classical survival probability is readily obtained as
\begin{equation}
P(t) = p_{1,1}(t) = e^{-t(\tilde E+\tilde\Gamma/2)} \frac{\cosh\big(\psi +
t \tilde V \cosh\psi\big)}{\cosh\psi},
\end{equation}
which for not too small $t$ gives rise to a simple exponential decay with
exponent $\lambda_+$. Taking $\tilde E = \tilde V$ and $\tilde \Gamma \ll
1$ one obtains $P(t) \sim e^{-t\tilde \Gamma/2}$, which is independent of
$\tilde E$ and $\tilde V$. We note the difference in the overall
exponential decay of the LvNE which was proportional to $\exp(-\Gamma t)$.
Comparing this to the PIMC results for strong couplings $\alpha$, we
obtain $\tilde\Gamma = 2\Gamma$. 

\section{Conclusions}

We have studied the trapping of excitations in dimers which are coupled to
a dissipative environment. By using the (approximative) Lindblad form of
the LvNE on one hand and the (numerically exact) PIMC calculations on the
other, we were able to specify the range of coupling parameters in which
both methods agree with each other. Moreover, matching the two approaches
by appropriately adjusting the coupling parameter $\alpha$ in the LvNE
allows to extrapolate the short-to-intermediate-time PIMC result to -- in
principle -- arbitrary long times. Since PIMC is numerically exact, the
combination with the LvNE is ideal for also studying large systems at long
times for a broad range of couplings.

\section*{Acknowledgements}
Support from the Deutsche Forschungsgemeinschaft (DFG) and the Fonds der
Chemischen Industrie is gratefully acknowledged. LM acknowledges computational
support from the Black Forest Grid of the computing center of the university of
Freiburg.


\end{document}